\documentclass[12pt,epsf,showkeys,nofootinbib]{article}

\usepackage{graphicx,float}
\usepackage{mathrsfs,array,multirow}
\usepackage{amstext}
\usepackage{subfigure}
\usepackage{bm,latexsym,amsmath,amsfonts,amssymb}
\usepackage[usenames,dvipsnames]{color}
\usepackage{color}
\usepackage[colorlinks=true,linkcolor=blue]{hyperref}
\usepackage{soul}
\usepackage{epsfig}

\newcommand{\be}[1]{\begin{equation} \label{#1}}
\newcommand{\ee}{\end{equation}}

\newcommand{\ba}{\begin{array}}
\newcommand{\ea}{\end{array}}

\newcommand{\bea}{\begin{eqnarray}}
\newcommand{\eea}{\end{eqnarray}}

\begin{document}
\begin{titlepage}
\vspace{.5in}
\begin{flushright}
\end{flushright}
\vspace{0.5cm}

\begin{center}
{\Large\bf Phase transition for a black hole with matter fields and the relation with the Lyapunov exponent }\\
\vspace{.4in}

  {$\mbox{Pakhlavon \,\, Yovkochev}^{\ddag}$}\footnote{\it email: yovqochevp@gmail.com},\,\,
  {$\mbox{Bobomurat J.\,\, Ahmedov}^{\natural\alpha\beta}$}\footnote{\it email: ahmedov@astrin.uz},\,\,
  {$\mbox{Bum-Hoon \,\, Lee}^{\S\dag\flat}$}\footnote{\it email: bhl@sogang.ac.kr},\,\,
  {$\mbox{Hocheol \,\, Lee}^{\sharp}$}\footnote{\it email: insaying@dongguk.edu},\,\,
  {$\mbox{Wonwoo \,\, Lee}^{\S}$}\footnote{\it email: warrior@sogang.ac.kr},\,\, \\

\vspace{.1in}

{\small \ddag \it Institute of Fundamental and Applied Research, National Research University TIIAME, Kori Niyoziy 39, Tashkent 100000, Uzbekistan}\\
{\small $\natural$ \it Institute of Theoretical Physics, National University of Uzbekistan, Tashkent 100174, Uzbekistan}\\
{\small $\alpha$ \it Institute for Advanced Studies, New Uzbekistan University, Movarounnahr str. 1, Tashkent 100000, Uzbekistan}\\
{\small \S \it Center for Quantum Spacetime, Sogang University, Seoul 04107, Korea}\\
{\small \dag \it Department of Physics, Sogang University, Seoul 04107, Korea}\\
{\small $\sharp$ \it Department of Physics, Dongguk University, Seoul 04620, Korea}\\
{\small $\beta$ \it School of Physics, Harbin Institute of Technology, Harbin 150001, China}\\
{\small $\flat$  \it  Department of Physics, Shanghai University, Shanghai, 200444, China}\\

\vspace{.2in}

\end{center}
\begin{center}
{\large\bf Abstract}
\end{center}
\begin{center}
\begin{minipage}{4.75in}

{\small \,\,\,\, We construct static black hole solutions coexisting with anisotropic matter in asymptotically (anti-)de Sitter spacetime and investigate how the matter field modifies their thermodynamic and dynamical properties. The anisotropic matter generates an exponentially decaying charge-like contribution to the metric, causing the geometry to interpolate between the Schwarzschild black hole in anti-de Sitter spacetime and the Reissner-Nordstr\"om black hole in anti-de Sitter spacetime. In anti-de Sitter black hole spacetime, we derive the complete thermodynamic description, including the Hawking temperature, heat capacity, Smarr relation, generalized first law, and Helmholtz free energy. The system exhibits a van der Waals-type small/large black hole phase transition with a critical point determined by the matter field parameters. Local thermodynamic stability is characterized by the heat capacity, whereas global stability is determined through the free energy. We further investigate unstable homoclinic orbits by evaluating the Lyapunov exponent associated with null geodesics. Our analysis reveals that different thermodynamic branches possess distinct dynamical instabilities, and that the thermodynamically preferred phase is accompanied by a smaller Lyapunov exponent. These results demonstrate that thermodynamic stability and geodesic instability are correlated because both originate from the same underlying spacetime geometry.
 }
\end{minipage}
\end{center}
\end{titlepage}

\newpage

\section{Introduction \label{sec1}}

\quad

The detection of gravitational waves coming from black hole collisions~\cite{LIGOScientific:2025rsn, LIGOScientific:2020iuh} and observations of black hole shadows~\cite{EventHorizonTelescope:2025vum, EventHorizonTelescope:2025whi, EventHorizonTelescope:2019dse} have further fueled interest in black holes in the Universe, serving as a test bed for various black hole geometries, including those coexisting with dark matter and dark energy. Concurrently, theoretical research on black hole thermodynamics continues to be conducted as one of numerous approaches to investigating
the quantum theory of gravitation.

Research exploring the analogy between a black hole system and thermodynamics~\cite{Bardeen:1973gs} has been extensively studied as a method to approximate quantum theory on a fixed black hole background~\cite{Bekenstein:1973ur, Hawking:1974rv, Hawking:1975vcx, Strominger:1996sh, Cohen:1998zx, Witten:2024upt}.
This research suggests that black holes should be regarded not merely as celestial bodies that engulf everything, but as thermodynamic systems endowed with temperature and entropy. In particular, the contributions by Bekenstein and Hawking initiated a scientific discourse by demonstrating that the area of a black hole's event horizon is proportional to its entropy~\cite{Bekenstein:1973ur, Hawking:1975vcx}. Furthermore, ongoing research efforts to develop the holographic principle~\cite{Maldacena:1997re,  Witten:1998qj} and understand the information paradox persist~\cite{Hawking:1976ra, Page:1993wv, Chen:2014jwq}.

As the relationship between black holes and thermodynamics developed~\cite{Smarr:1972kt, Hawking:1982dh, Wald:1993nt, Jacobson:1993vj}, research emerged into what constitutes a thermodynamically stable state for a black hole. The concept of local stability is examined through the sign of specific heat (or heat capacity)~\cite{Hawking:1976de, Davies:1977bgr}.
Various black hole systems are being analyzed to ascertain their global stability~\cite{Braden:1990hw, Peca:1998cs, Chamblin:1999tk, Cai:2001dz, Cho:2002hq, Cai:2007wz, Myung:2008eb, Khimphun:2016gsn, Park:2016wch, Su:2019gby, Eom:2022nwc, Ge:2023ghy, Lin:2024gip, Jeon:2024yey, Gohain:2024piy, Braga:2025jji, Guo:2025flg, Lee:2026uvo}.
In \cite{Hawking:1982dh}, it depends on the sign of the free energy difference between the black hole system in an anti-de Sitter spacetime
and a system consisting of thermal particles in the anti-de Sitter one.
Global stability may correspond to determining which of the compared geometries is more probable.
It is important to note that this sensitivity to initial conditions may not be related to either of the two opposing behaviors: the particle's entire trajectory being integrable or exhibiting global chaotic behavior.

Meanwhile, when analyzing particle motion in the spacetime geometry around a black hole, homoclinic orbits may exist as purely curvature effects of spacetime~\cite{Levin:2008yp, Perez-Giz:2008ajn, Dariescu:2023twk}, absent in the Keplerian orbits of the Newtonian theory of gravitation. There is research that uses the Lyapunov exponent as an indicator to check the sensitivity of initial conditions for these orbits,
to examine the particle's local chaotic behavior~\cite{Cardoso:2008bp, Zhao:2018wkl, Lei:2020clg, Addazi:2021pty, Gwak:2022xje, Gao:2022ybw, Xie:2023tjc, Giataganas:2024hil, Lei:2024qpu, Singh:2024qfw, Targema:2025qws, Lee:2025vih, Lungu:2025pgk, Lee:2025ias, Targema:2025opv}.
It is important to note that this sensitivity to initial conditions may not be related to either of the two opposing behaviors~\cite{Jeong:2023hom}: the particle's entire trajectory being integrable, or exhibiting global chaotic behavior~\cite{Dutta:2024rta, Ciou:2025ygb, Das:2025vja, An:2025xmb, Dalui:2026msw, Xu:2026tgb}.

Recent research has investigated the correlation between global stability and Lyapunov exponent analysis in the context of black hole free energy analysis. In the context of free energy analysis of black hole systems, the presence of multiple values of free energy is observed. It is demonstrated that the system with the lowest free energy manifests global stability. Furthermore, the Lyapunov exponent values across multiple black hole systems can vary significantly. At the temperature at which a phase transition occurs, the system with the lowest Lyapunov exponent value could possess global stability~\cite{Guo:2022kio, Yang:2023hci, Shukla:2024tkw, Gogoi:2024akv, Gao:2025rep}.

We construct a black hole geometry that coexists with a specific type of anisotropic matter~\cite{Kim:2025sdj} in the (anti)-de Sitter spacetime. As an applied study of this black hole geometry, we investigate the black hole thermodynamics and the phase transition in anti-de Sitter spacetime while examining the effects of matter fields~\cite{Kim:2025sdj, Kiselev:2002dx, Cho:2017nhx, Toshmatov:2015npp, Kim:2019hfp, Li:2025glq, UktamjonUktamov:2025emm, Lee:2026cit, Kim:2026gog}.

This paper is organized as follows:  In Sec.~\ref{sec2}, we construct a black hole that coexists with both anisotropic matter and the cosmological constant. Although our construction allows for both positive and negative cosmological constants, the present work is concerned with black holes in the presence of a negative cosmological constant. Consequently, we investigate the horizon structure for such cases. In Sec.~\ref{sec3}, we investigate the thermodynamic stability of black holes. We look at the temperature of the black hole, the heat capacity(specific heat), and the free energy. In Sec.~\ref{sec4}, we examine the Lyapunov exponent and compare it with the free energy obtained in Sec.~\ref{sec3} to analyze the correlation between this quantity and the thermodynamic global stability of black holes.

\section{Black hole with a matter field in (anti-)de Sitter spacetime \label{sec2}}
\quad

In this section, we construct a black hole with anisotropic matter in (anti-)de Sitter spacetime.
As the solution for a black hole with fluid matter is already known (see Ref.~\cite{Kim:2025sdj}),
our aim is to construct a black hole solution incorporating an additional cosmological constant based on this solution.
To do this, we consider the action~\footnote{Due to the absence of a recognized action for such fluid matter, it is difficult to show the corresponding explicit action; as an alternative, one may begin with an action that includes a nonlinear electrodynamics term~\cite{Kim:2019hfp, Lee:2026cit}.} with the cosmological constant
\begin{equation}
\label{action}
I=\int d^4x \sqrt{-g}\Big[\frac{R-2\Lambda}{2\kappa}+{\cal L}_{\rm am}\Big] + I_{\rm b} \,,
\end{equation}
where $\kappa=8\pi G$, $\Lambda$ is the positive value for the de Sitter spacetime,  while it takes the negative value for the anti- de Sitter spacetime, ${\cal L}_{\rm am}$ describes an effective anisotropic matter field and $I_{\rm b}$ is the boundary term.

From the variation, we obtain the Einstein equations
\begin{equation}
\label{einsteineq}
G_{\mu\nu}=R_{\mu\nu}-\frac{1}{2}R g_{\mu\nu}=\kappa T_{\mu\nu} \,,
\end{equation}
where the stress-energy tensor takes the form
\begin{eqnarray}
\label{stente}
T^{\mu\nu}&=& T^{\mu\nu}_{\rm am} - \frac{\Lambda g^{\mu\nu}}{\kappa} \,,
\end{eqnarray}
where $T^{\mu\nu}_{\rm am}$ can be rewritten as
$T^{\mu}_{\nu {\rm am} }={\rm diag}(-\varepsilon_{\rm am}, p_{r {\rm am}},  p_{t {\rm am}}, p_{t {\rm am}})
={\rm diag}(-\varepsilon_{\rm am}, - \varepsilon_{\rm am}, w \varepsilon_{\rm am}, w \varepsilon_{\rm am})$,
in which $w= \left(1+ \frac{v_c r^2}{2} \right)$ as shown in \cite{Kim:2025sdj}. Then, the energy density and pressures are given by
\begin{eqnarray}
\varepsilon&=& \frac{\Lambda}{\kappa} +\varepsilon_{\rm am} = \frac{\Lambda}{\kappa} + \frac{v_2 e^{-v_c r^2/2}}{\kappa r^4}  \,, \nonumber \\
p_r &=& - \varepsilon = -\frac{\Lambda}{\kappa}  -  \varepsilon_{\rm am}\,, \\
p_{\theta} &=& p_{\phi} =  -\frac{\Lambda}{\kappa}  + \left(1+ \frac{v_c r^2}{2} \right) \varepsilon_{\rm am}\,. \nonumber
\label{edenpre}
\end{eqnarray}
Our construction of a black hole with a cosmological constant is not affected by its sign.
However, this study would focus on the case of a negative cosmological constant.
We now mention the energy condition. Due to the negative cosmological constant, the weak energy condition could be violated for $v_2>0$ and $v_c>0$, while the null energy condition is satisfied as $p_r+\varepsilon =0$.
Our main research interest focuses on black hole phase transitions in the anti-de Sitter spacetime.
Since the violation of the weak energy condition is induced by the negative vacuum of the anti-de Sitter spacetime
rather than the matter fields, the situation remains safe owing to the fulfillment of the null energy condition.
However, most of the energy conditions are satisfied when the negative cosmological constant vanishes.
\begin{figure}[H]
    \centering
    \subfigure[Energy density with varying $v_c$]{\includegraphics[width=0.49\textwidth]{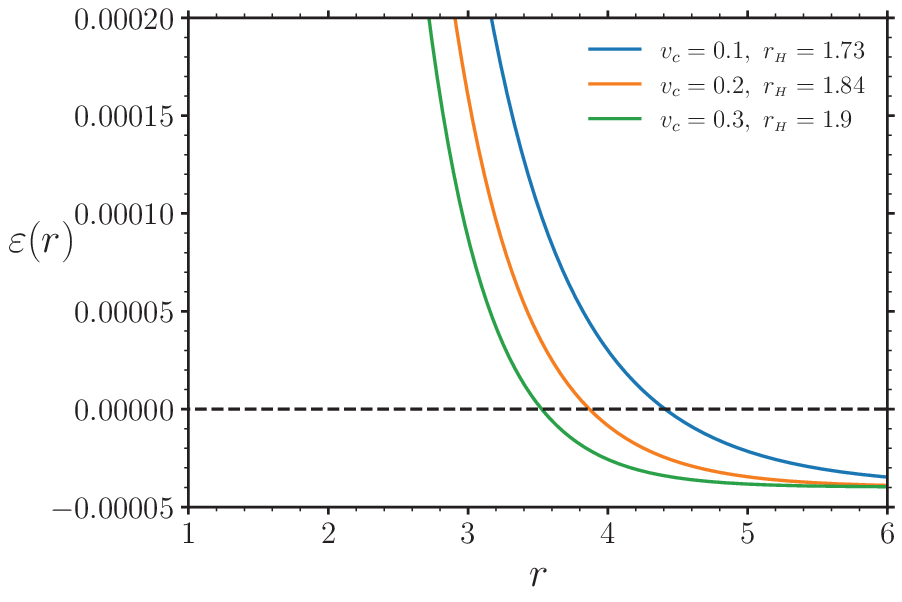}}
    \subfigure[Tangential pressure with varying $v_c$]{\includegraphics[width=0.49\textwidth]{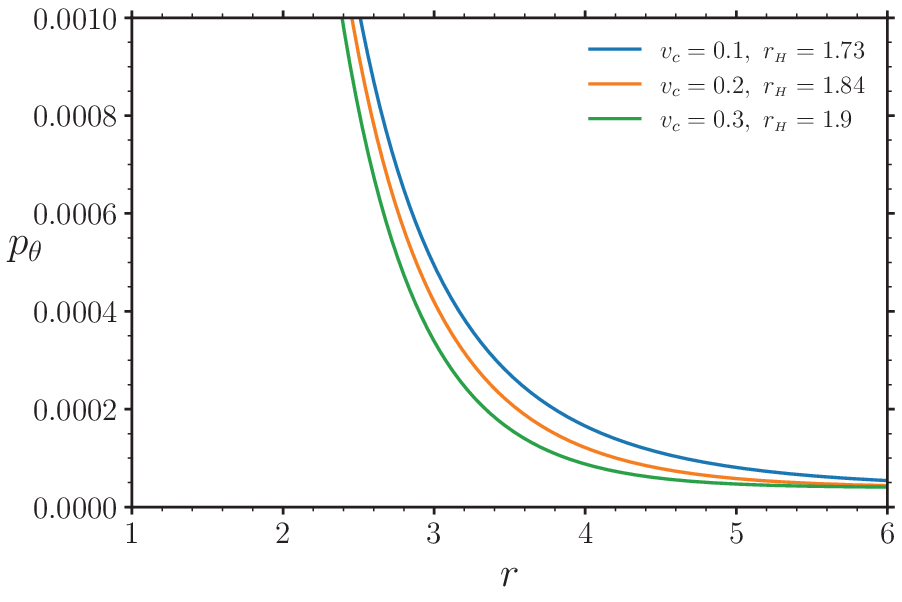}}
    \caption{Energy density $\varepsilon(r)$ and pressure $p_{\theta}(r)$ as a function of $r$ in the AdS spacetime.}
    \label{fig:energyden}
\end{figure}
Figure \ref{fig:energyden} shows the graph of the energy density and the tangential pressure as a function of $r$
with $v_2=1.0$, $M=1$, and $\Lambda=-0.001$.
The blue curve represents the plot when $v_c$ is $0.1$, the orange curve when $v_c$ is $0.2$,
and the green curve when $v_c$ is $0.3$, respectively.
For a given value of $v_2$, the energy density and the tangential pressure increase as the value of $v_c$ decreases.
Although the numbers for each physical quantity are not explicitly defined as dimensionless values,
they should be considered as such hereafter.

We take the metric ansatz to be
\begin{equation}
\label{stmetric}
ds^2=  - f(r) dt^2 +  \frac{1}{f(r)} dr^2 + r^2 d\Sigma^2_k  \,,
\end{equation}
where the metric function is given by
\begin{equation}
\label{metricfunc}
f(r) = k - \frac{2M}{r} + G(r) - \frac{\Lambda}{3}r^2 \,,
\end{equation}
where $k=1$ for the de Sitter spacetime, and $k=1$, $0$, or$-1$ for the anti-de Sitter spacetime,
in which $k = 1$, $0$ or $-1$ correspond to a surface of positive, zero, or negative Gaussian curvature, respectively~~\cite{Mann:1996gj}. Therefore, $d\Sigma^2_k$ indicates
$d\Sigma^2_1 =d\theta^2 + \sin^2\theta d\phi^2$ for spherically  symmetric ones with $k=1$,
$d\Sigma^2_0 =dx^2 + dy^2$  for plane symmetric ones with $k=0$,
and $d\Sigma^2_{-1} = d\psi^2 + \sinh^2\psi d\phi^2$ for  hyperbolic symmetric ones with $k=-1$, respectively.

The function $G(r)$ is given by
\begin{equation}
G(r) = \frac{v_2}{r^2}\left(e^{-\tfrac{v_c r^2}{2}} + r \sqrt{\frac{\pi v_c}{2}}\left[\mathrm{erf}\!\left(r\sqrt{\tfrac{v_c}{2}}\right) - 1\right]\right)\,.
\end{equation}
It reduces to the black hole obtained in \cite{Kim:2025sdj} when vanishing $\Lambda$, while it reduces to the Schwarzschild anti-de Sitter black hole when vanishing $v_2$~\cite{Kottler:1918cxc, Mann:1996gj}.
Hereafter, the Schwarzschild anti-de Sitter black hole will be denoted SAdS.
It reduces to the Reissner-Nordstr\"om anti-de Sitter black hole when non-vanishing $v_2$ and vanishing $v_c$,
while this one possesses a charged-like term that is analogous to a charged black hole, but decreases exponentially with respect to $r$ when both non-vanishing $v_2$ and $v_c$. Hereafter, the Reissner-Nordstr\"om anti-de Sitter black hole will be denoted RNAdS.

We now analyze the structure of the horizon of black hole geometries that we have constructed.
In this geometry, the presence of additional fluid matter gives rise to two black hole horizons at most, such as $r_+$ and $r_-$.
From now on, we use $r_H$ to represent $r_+$.
The horizon satisfies the $r_H = 0$ condition in Eq.~\eqref{metricfunc}.
If the following conditions are satisfied, an extremal black hole could exist:
\begin{equation}
\label{conextbh}
f(r)|_{r_H} = 0\,, f'(r)|_{r_H} =0\,,  f''(r)|_{r_H} > 0 \,.
\end{equation}

\begin{figure}[H]
    \centering
    \subfigure[The cases with $k=1$ and varying $v_2$]{\includegraphics[width=0.41\textwidth]{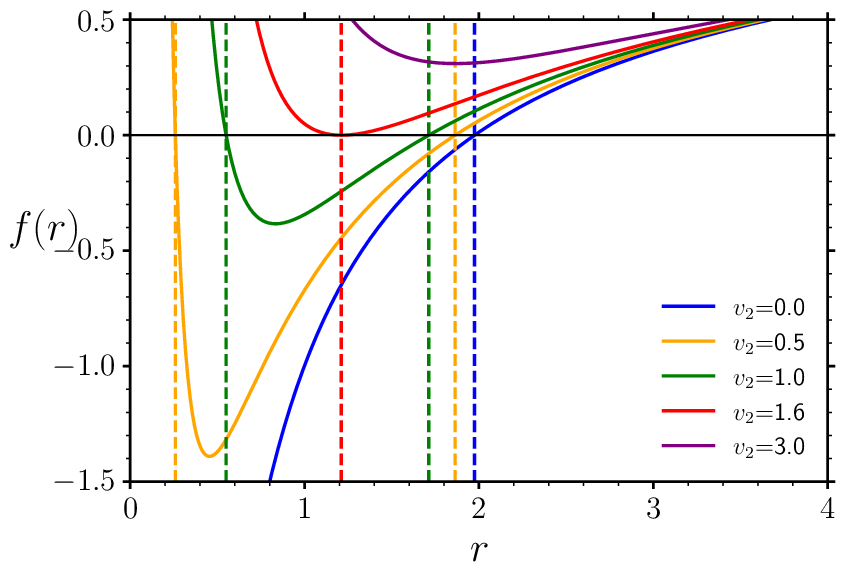}}
    \subfigure[The cases with $k=0$ and varying $v_2$]{\includegraphics[width=0.41\textwidth]{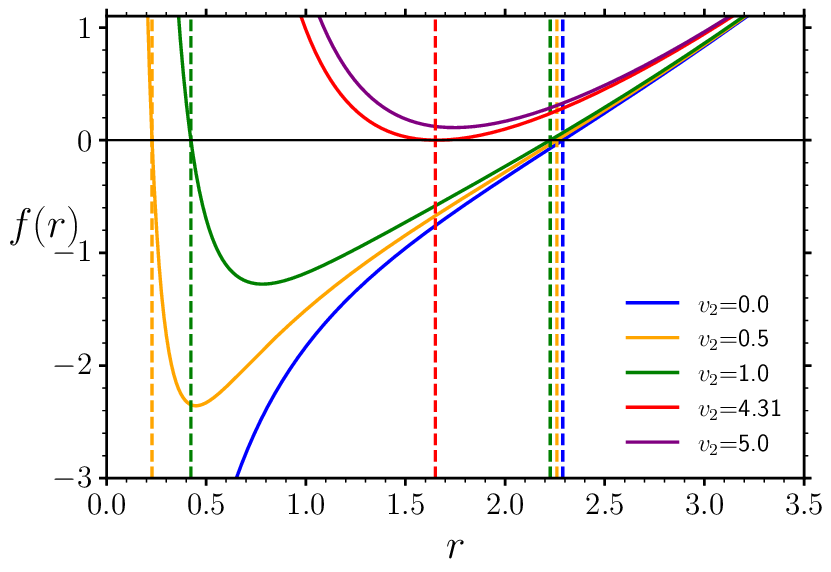}} \\
    \subfigure[The cases with $k=-1$ and varying $v_2$]{\includegraphics[width=0.41\textwidth]{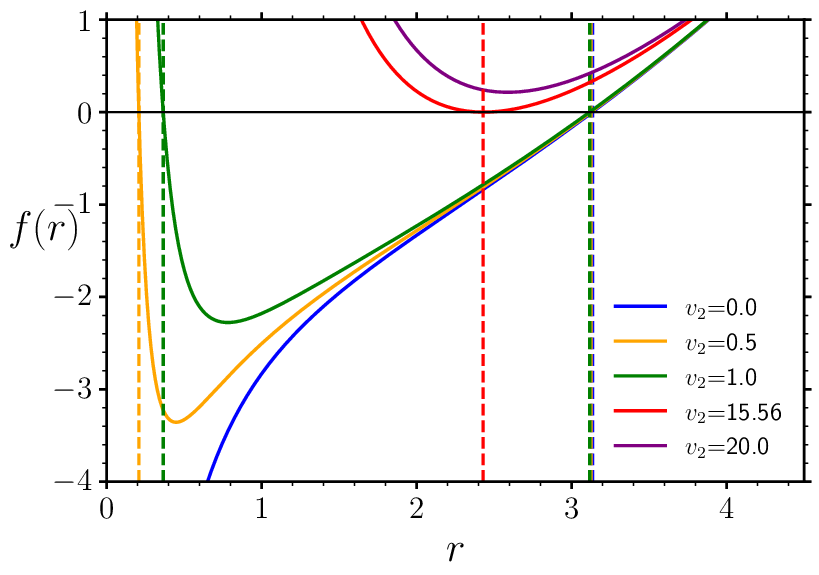}}
    \subfigure[Division region based on the number of black hole horizons with $k=1$ within the parameter space]
    {\includegraphics[width=0.41\textwidth]{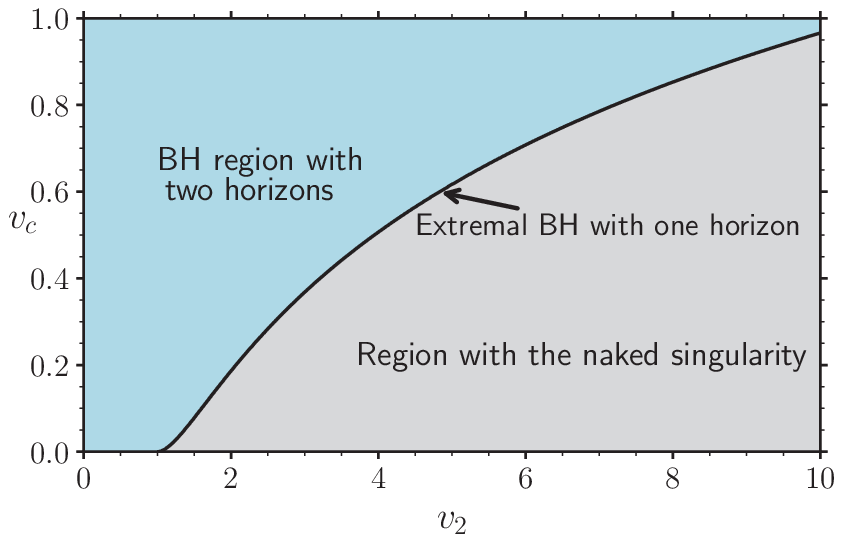}}
    \caption{Metric function $f(r)$ as a function of $r$. }
    \label{fig:horizan}
\end{figure}
Figure \ref{fig:horizan} shows the plots of the metric function $f(r)$ as a function of $r$ for given parameter values,
$v_c=0.1$ and $M=1$.
This shows the horizon structures of these black holes.
Figure $(a)$ shows the case where $k=1$ with $\Lambda=-0.001$,
while $(b)$ and $(c)$ show the cases where $k=0$ and $k=-1$ with $\Lambda=-0.5$, respectively.
The vertical dotted lines of each color indicate the locations of the black hole horizons.
The blue curve shows the SAdS case with the vanishing $v_2$ as reference.
The SAdS case has the largest horizon.
As expected, as the value of $v_2$ increases for a fixed value of $v_c$,
the positions of the two horizons come together to form a single point.
This is what we call an extremal black hole case.
If $v_2$ increases any further, it becomes a case of a naked singularity.
The red curves in $(a)$, $(b)$, and $(c)$ show the case of extremal black holes with a single degenerate horizon.
Figure $(d)$ shows the division region plot, which is based on the number of black hole horizons within the parameter space.

\section{Thermodynamic properties and stabilities of black holes \label{sec3}}

\quad
In this section, we present the key thermodynamic characteristics of the black hole, including the Hawking temperature, entropy, heat capacity, and free energy. These quantities are crucial for analyzing the thermal behavior and phase structure of the black hole in the AdS spacetime.

\subsection{Black hole temperature \label{sec3.1}}
\quad

The Hawking temperature~\cite{Hawking:1974rv, Hawking:1975vcx} is given by
\begin{equation}
\label{Tem}
T_H =  \frac{k r^2_H - v_2 e^{-v_c r^2_H /2 } -\Lambda r^4_H }{4\pi r^3_H}\,,
\end{equation}
where $M$ is not explicitly written using the horizon information, $f(r_H)=0$.
\begin{figure}[H]
    \centering
    \subfigure[$T_H$ with $v_c=0.5$ and varying $v_2$]{\includegraphics[width=0.49\textwidth]{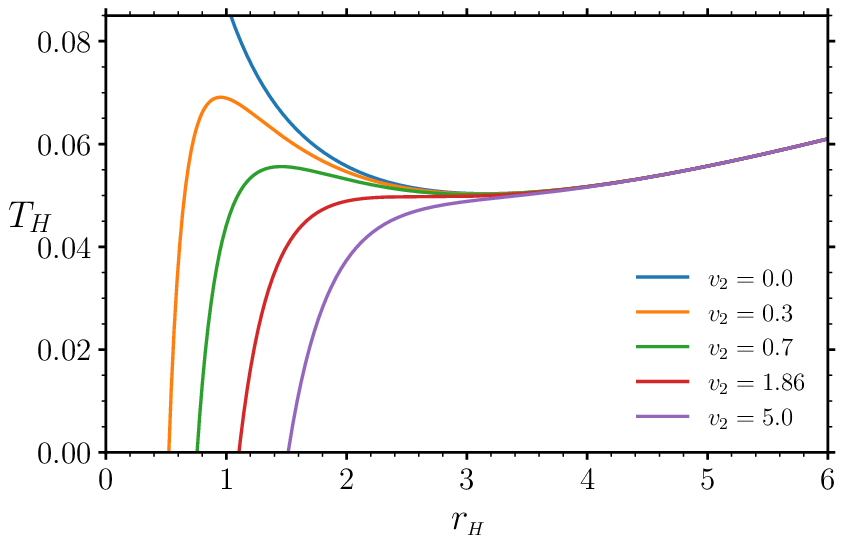}}
    \subfigure[$T_H$ with $v_2=2.0$ and varying $v_c$]{\includegraphics[width=0.49\textwidth]{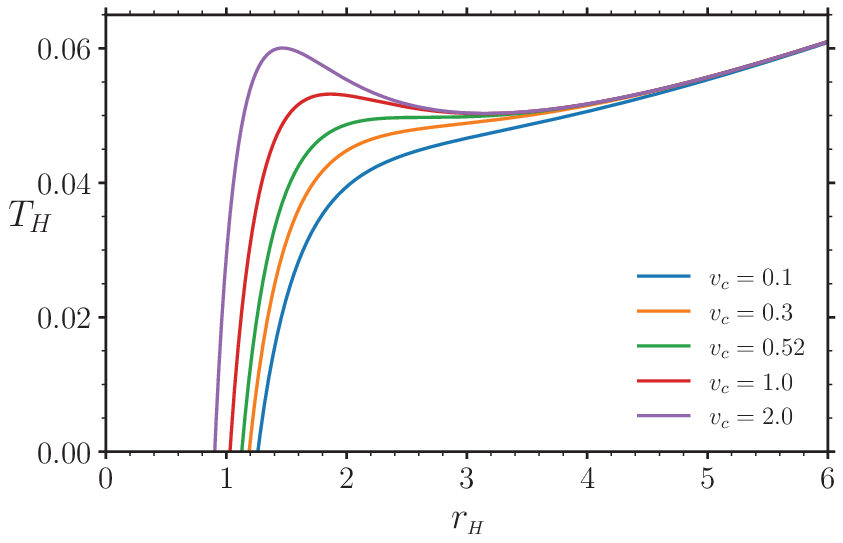}}
    \caption{Black hole temperature $T_H$ versus horizon radius $r_H$. }
    \label{fig:temperaturevc}
\end{figure}
Figure~\ref{fig:temperaturevc} shows plots of the black hole temperature $T_H$ as a function of $r_H$
with $k=1$, $\Lambda = -0.1$ and varying parameter values.
Figure $(a)$ shows the black hole temperature as a function of $r_H$ for different $v_2$ values,
where $v_c$ is $0.5$. When $v_2$ is $0$, it corresponds to the SAdS case, which is the reference case.
The orange curve shows $v_2 = 0.3$, the green curve $v_2 = 0.7$, the red curve $v_2 = 1.86$ as the critical case,
and the purple curve $v_2 = 5.0$, respectively.
A critical case occurs when the point at which the first derivative of the temperature is zero, $dT_H/d r_H =0$,
coincides with the point at which the second derivative is also zero, $d^2 T_H/d{r_H}^2$.
Figure $(b)$ shows the black hole temperature as a function of $r_H$ for different $v_c$ values, where $v_2$ is $2.0$.
The purple curve shows $v_c = 2.0$, the red curve $v_c = 1.0$, the green curve $v_c = 0.52$ as the critical case,
the orange curve $v_c = 0.3$, and the blue curve $v_c = 0.1$, respectively.

We now analyze the temperature curve of the black hole.
We divide it into three regions and use the SAdS case as a reference.
These are the matter-dominated, mass-dominated, and AdS-dominated regions.
We then revisit the critical case.

\begin{itemize}
\item Small $r_H$ (matter-dominated region)
\label{temp01}
\end{itemize}
We analyze three different regions: the small, intermediate, and large horizon radius.
First, we expand the exponential in Eq.~\eqref{Tem} for $ \sqrt{\frac{v_c}{2}} r_H  \ll1 $ to yield
\begin{equation}
T_H(r_H) \simeq -\frac{v_2}{4\pi r_H^3}
+ \frac{k+\tfrac{1}{2}v_2 v_c}{4\pi r_H}
+ \frac{-\Lambda - \tfrac{1}{8}v_2 v_c^2}{4\pi}\, r_H + \mathcal{O}(r_H^3) \,.
\end{equation}
The derivative then behaves as
\begin{equation}
\label{srtepde}
\frac{dT_H}{dr_H} \simeq \frac{1}{4\pi}\left(\frac{3v_2}{r_H^4} - \frac{k+\tfrac{1}{2}v_2 v_c}{r_H^2} - \Lambda \right) + \mathcal{O}(r_H^0) \,.
\end{equation}
As $r_H \to 0$ the $+3v_2/(4\pi r_H^4)$ term dominates, so $dT_H/dr_H>0$.
Therefore, the very small $r_H$ has a positive slope, and the matter term $v_2$ supports the positive slope, which could show  locally stable behavior, as shown in Fig.~\ref{fig:temperaturevc}.

\begin{itemize}
\item Intermediate $(r_H)$ (mass-dominated region)
\label{temp02}
\end{itemize}
For $v_c^{-1/2} \ll r_H \ll L_\Lambda$ (where $L_\Lambda\sim|\Lambda|^{-1/2}$ is the AdS-dominated length scale),
the exponential term is exponentially suppressed, and the$\Lambda$-term is still small. Thus,
\begin{equation}
\label{mrtepde}
T_H(r_H)\simeq \frac{k}{4\pi r_H},\qquad
\frac{dT_H}{dr_H}\simeq -\frac{k}{4\pi r_H^2}<0 \,.
\end{equation}
The intermediate region in Fig.~\ref{fig:temperaturevc} follows the Schwarzschild-like behavior with negative slope which could show  locally unstable behavior.

\begin{itemize}
\item Large $r_H$ (AdS-dominated region)
\label{temp03}
\end{itemize}
For large $r_H$, the exponential vanishes and the $\Lambda$-term dominates:
\begin{equation}
T_H(r_H) \simeq \frac{1}{4\pi}\Big(\frac{k}{r_H}-\Lambda r_H\Big) \,.
\end{equation}
Hence,
\begin{equation}
\label{lrtepde}
\frac{dT_H}{dr_H} \simeq \frac{1}{4\pi}\Big(-\frac{k}{r_H^2}-\Lambda\Big) \,.
\end{equation}
For AdS $( \Lambda < 0 )$ the term $-\Lambda > 0 $ eventually dominates, thus the slope becomes positive at sufficiently large $r_H$.
This corresponds to the positive-slope on the right which could show locally stable behavior, as shown in Fig.~\ref{fig:temperaturevc}.

\begin{itemize}
\item Critical line in the $(v_2,v_c)$ parameter space
\label{temp04}
\end{itemize}
The thermodynamic structure of the black hole is strongly dependent on the parameters of anisotropic matter  $(v_2, v_c)$.
To characterize the onset of the multi-branch temperature behavior, we determine the critical values of the parameters
for which the two extrema of the temperature curve merge into a single inflection point.
This occurs when the conditions
\begin{equation}
\label{temde}
\frac{dT_H}{dr_H}=0, \qquad \frac{d^2T_H}{dr_H^2}=0
\end{equation}
are satisfied simultaneously.
\begin{figure}[H]
    \centering
    \includegraphics[width=0.5\linewidth]{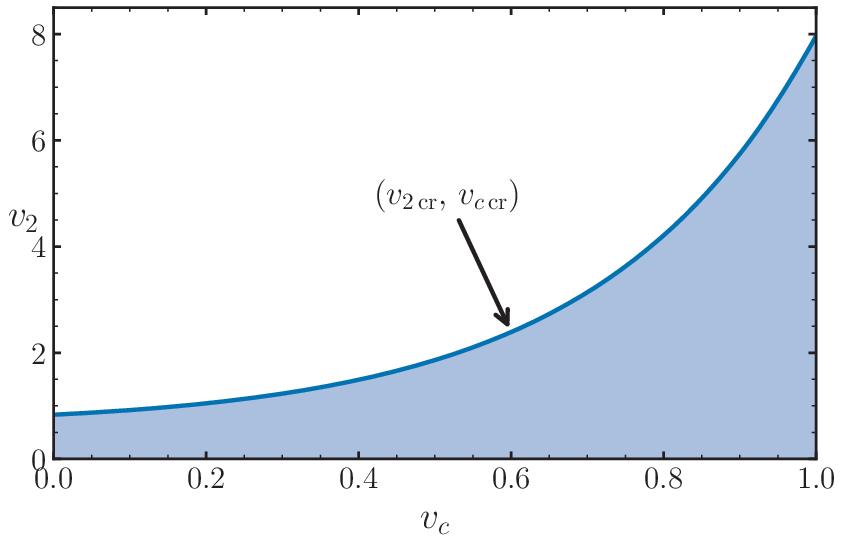}
    \caption{The critical line in $(v_2,v_c)$ parameter space separating regions with and without a triple-branch black hole structure.}
    \label{fig:critical}
\end{figure}
Figure \ref{fig:critical} shows the plot of the critical line in the parameter space $(v_2,v_c)$ separating regions with and without a triple-branch temperature structure. Above this line, $T_H(r_H)$ is monotonic;
below it, the system exhibits a van der Waals-like phase structure.
Solving Eq.~\eqref{temde} for each different $v_c$ value yields the critical relationship $v_{2,\rm{cr}}(v_c)$,
which distinguishes two qualitatively different thermodynamic regions and provides the resulting critical curve.
Above the critical line, the temperature $T_H(r_H)$ varies monotonically with the radius of the horizon,
and therefore the black hole possesses a single thermodynamic branch without phase transitions,
which can be seen in Fig.~\ref{fig:temperaturevc}.
In contrast, below this curve the temperature develops a local maximum and a local minimum,
producing a characteristic triple-branch structure.
In this regime the system admits three black hole solutions at a given temperature:
a small black hole branch, an intermediate black hole branch, and a large black hole branch.

The entropy of the black hole is proportional to the area of the event horizon~\cite{Bekenstein:1973ur, Hawking:1975vcx}:
\begin{equation}
\label{entropy}
S = \frac{A}{4} = \pi r_H^2 \,.
\end{equation}
This entropy would be a measure of the information lost due to collapse.
It will be used in computing other thermodynamic quantities, such as heat capacity and free energy.

\subsection{Heat Capacity \label{sec3.2}}

\quad
The local stability of a system would be determined by its heat capacity(specific heat)~\cite{Hawking:1976de, Davies:1977bgr}.
A positive value indicates local thermodynamic stability, while a negative value signals instability.
In the AdS spacetime, large black holes can exhibit a positive heat capacity, leading to a stable thermal equilibrium.

The heat capacity characterizes how the black hole's temperature changes with respect to its energy or, alternatively, by using the entropy:
\begin{equation}
C = T_H \left( \frac{\partial S}{\partial T_H} \right) = \frac{2 \pi r_H^2 \left(r_H^2 e^{\frac{r_H^2 v_c}{2}} \left(\Lambda  r_H^2-k\right)+v_2\right)}{r_H^2 e^{\frac{r_H^2 v_c}{2}} \left(k+\Lambda r_H^2\right)-v_2 \left(r_H^2 v_c+3\right)} \,,
\end{equation}
where $v_2$, $v_c$, $k$, and $\Lambda$ are fixed for the partial derivative.

\begin{figure}[H]
    \centering
    \subfigure[$C_{r_H}$ with $v_c=0.5$ and varying $v_2$]{\includegraphics[width=0.49\textwidth]{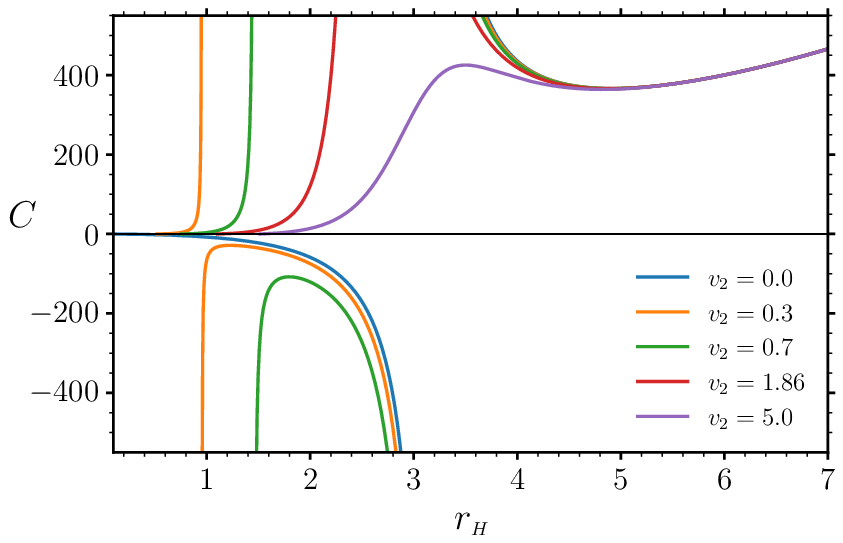}}
    \subfigure[$C_{r_H}$ with $v_2=2.0$ and varying $v_c$]{\includegraphics[width=0.49\textwidth]{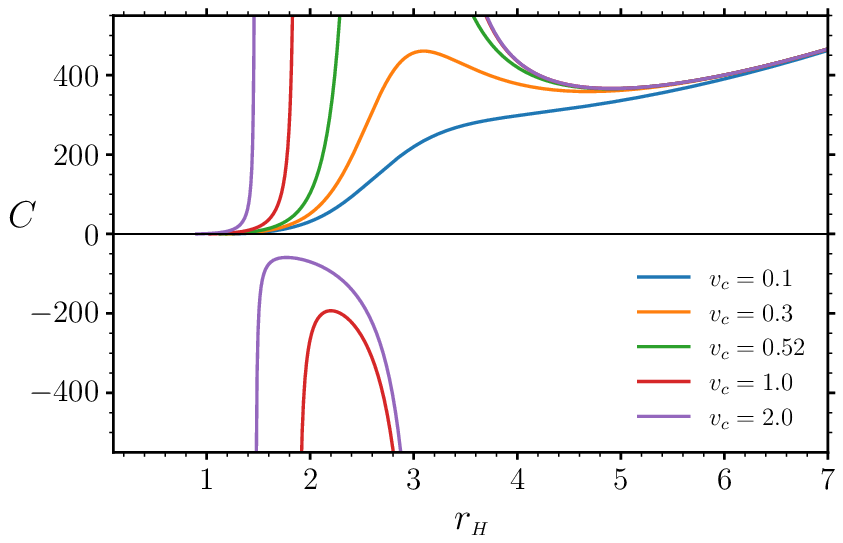}}
    \caption{Heat capacity $C_{r_H}$ versus horizon radius $r_H$ for parameter values according to Fig.~\ref{fig:temperaturevc}. }
    \label{fig:heatcapacity}
\end{figure}
Figure \ref{fig:heatcapacity} shows the graph of the heat capacity(specific heat) as a function of $r_H$ with $k=1$ and $\Lambda = -0.1$.
As shown in this figure, we plan to analyze $C_{r_H}$ by dividing it into three regions as; locally stable region ($C_{r_H}>0$) around  small $r_H$, an intermediate unstable region ($C_{r_H} < 0$), and the locally stable region ($C_{r_H}>0$) around large $r_H$, according to the approach of analyzing the temperature in three regions.
The tuning of $v_2$ and $v_c$ shifts these points, and one could remove the intermediate unstable region at critical parameter values (e.g. $v_2 \approx 1.86$ for $v_c=0.5)$, $v_c \approx 0.52$ for $v_2=2.0$.
Figure $(a)$ shows the heat capacity as a function of $r_H$ for different values of $v_2$, where $v_c$ is $0.5$.
The blue curve shows the SAdS case, which is used as a reference.
In this case, the region is divided into two areas: a locally unstable region dominated by mass
and a locally stable region dominated by AdS where $r_H$ is large.
For curves of different colors, each is the case adopted in Fig.~\ref{fig:temperaturevc}$(a)$.
Figure $(b)$ shows the heat capacity as a function of $r_H$ for different values of $v_c$, where $v_2$ is $2.0$.

\begin{itemize}
\item Small $r_H$ (matter-dominated region)
\label{cap01}
\end{itemize}
As shown in Eq.~\eqref{srtepde} of the previous subsection, the rate of change of the black hole temperature
with respect to $r_H$ is greater than zero in the region where $r_H$ is small.
This suggests the following implications:
\begin{equation}
\frac{\partial T_H}{\partial r_H} > 0 \qquad \Rightarrow \qquad C_{r_H} > 0 \,.
\end{equation}
Thus, unlike the SAdS case, the black hole is locally stable in the small-$r_H$ region.

\begin{itemize}
\item Intermediate $(r_H)$ (mass-dominated region)
\label{cap02}
\end{itemize}
As shown in Eq.~\eqref{mrtepde} in the previous subsection, in the intermediate $r_H$ region
(the region dominated by the black hole's mass),
the rate of change of the temperature of the black hole is less than zero with respect to $r_H$.
This implies the following:
\begin{equation}
\frac{\partial T}{\partial r_H} < 0  \qquad \Rightarrow \qquad C_{r_H} < 0  \,.
\end{equation}
This corresponds to the negative heat capacity, resulting in the local instability.

\begin{itemize}
\item Large $r_H$ (AdS-dominated region)
\label{cap03}
\end{itemize}
As shown in Eq.~\eqref{lrtepde} in the previous subsection, the influence of $\Lambda$ is greatest in regions where $r_H$ is large.
Consequently, the rate of change of the black hole temperature becomes positive
and the sign of $C_{r_H}$ becomes positive:
\begin{equation}
\frac{\partial T}{\partial r_H} > 0 \qquad \Rightarrow \qquad C_{r_H} > 0 \,.
\end{equation}
This leads to the existence of a locally stable black hole region once again.

\begin{itemize}
\item Critical line in the $(v_2,v_c)$ parameter space
\label{cap04}
\end{itemize}
The boundaries between the three regimes are divided by a divergence of the heat capacity, respectively.
Varying the parameters $(v_2 , v_c)$ modifies the stability structure of the heat capacity.

For sufficiently large $v_2$ or appropriately tuned $v_c$, the intermediate unstable region shrinks and may disappear entirely.
In this case, the divergence of $C_{r_H}$ occurs in the positive $C_{r_H}$ region, thus
the heat capacity becomes positive for all $r_H$ values.
This signals the presence of a thermodynamic critical point,
beyond which the small to large black hole phase transition ceases to exist (see Sec.~\ref{sec3.4}).
Representative critical values extracted from the plots include
\begin{equation}
v_2^{\mathrm{crit}} \simeq 1.86 \quad (v_c = 0.5)\,,
\qquad
v_c^{\mathrm{crit}} \simeq 0.52 \quad (v_2 = 2.0) \,.  \nonumber
\end{equation}
In the parameter region where the value exceeds this critical point,
the heat capacity is positive everywhere.
This makes the black hole with the parameter values locally stable for all $r_H$ regions (see Fig.~\ref{fig:heatcapacity}).

\subsection{Smarr relation and the first law of black hole thermodynamics \label{sec3.3}}

\quad
Even though the anisotropic matter field has been introduced, the exact form of the first law of black hole thermodynamics remains unknown.
For this reason, we first derive Smarr's mass relation using information that defines the event horizon of the black hole, $f(r_H)=0$. The Smarr relation~\cite{Smarr:1972kt} is given by
\begin{eqnarray}
M && = \frac{k r_H}{2} + \frac{v_2}{2r_H} \left[ e^{-v_c r_H^2/2} + r \sqrt{\frac{\pi v_c}{2}} \left( {\rm erf} \left(r_H \sqrt{\frac{v_c}{2}} \right) -1  \right)  \right] - \frac{\Lambda}{6} r^3_H \nonumber \\
&& = 2T_H S + \Phi_1 D_1 + \Phi_2 D_2 -2 pV\,,
\end{eqnarray}
where $\Phi_1 = \frac{\sqrt{v_2}}{r_H} e^{-v_c r_H^2/2}$, $D_1=\sqrt{v_2}$, $\Phi_2=\sqrt{v_2} v^{1/4}_c \sqrt{\frac{\pi}{8}} \left( {\rm erf}\left( \sqrt{\frac{v_c}{2}} r_H \right) -1  \right)$, $D_2=\sqrt{v_2} v^{1/4}_c$, $V=\frac{4}{3} \pi r^3_H$, $p=-\frac{\Lambda}{8\pi}$~\cite{Dolan:2010ha}.
The mass $M$ corresponds to the internal energy~\cite{Kastor:2009wy}.
We then differentiate this relation to derive the form of the first law.
The first law of black hole thermodynamics is given by
\begin{equation}
\delta M = T_H \delta S + \Phi_1 \delta D_1 + 2 \Phi_2 \delta D_2 -p \delta V \,.
\end{equation}

\subsection{Free Energy \label{sec3.4}}

\quad

We now analyze the free energy of the black hole system and the phase transition of the black hole~\cite{Hawking:1982dh}.
Using Eqs.~\eqref{Tem} and \eqref{entropy}, the Helmholtz free energy, $F = M - T_H S$, is given by
\begin{equation}
\label{freeeng}
F = \frac{1}{12}\left[ 3k r_H +\frac{9v_2 e^{-v_c r^2_H/2} }{r_H} + \Lambda r^3_H -3v_2 \sqrt{2\pi v_c} \left(1- {\rm erf} \left( r_H \sqrt{\frac{v_c}{2}} \right)  \right)   \right] \,.
\end{equation}
In the AdS spacetime with thermal particles, the mass, entropy, and free energy are all zero.

\begin{figure}[H]
    \centering
    \subfigure[$v_c=0.5$]{\includegraphics[width=0.49\textwidth]{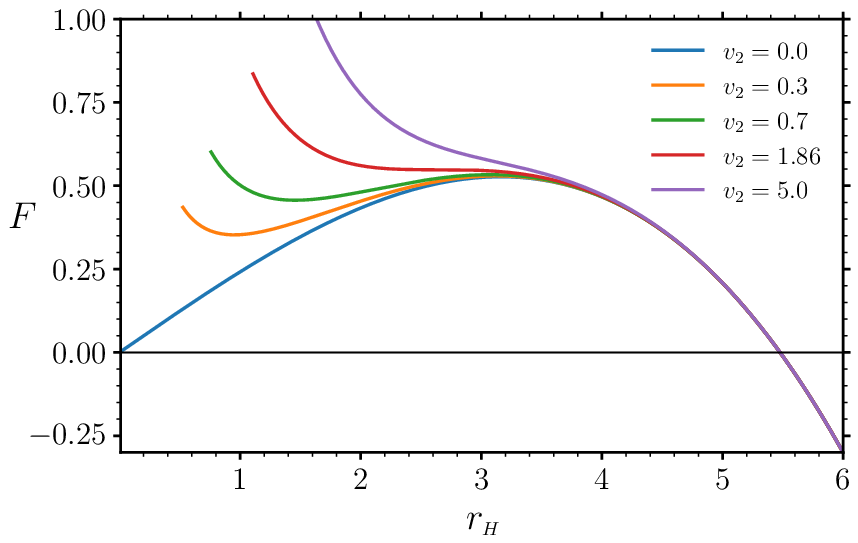}}
    \subfigure[$v_2=2.0$]{\includegraphics[width=0.49\textwidth]{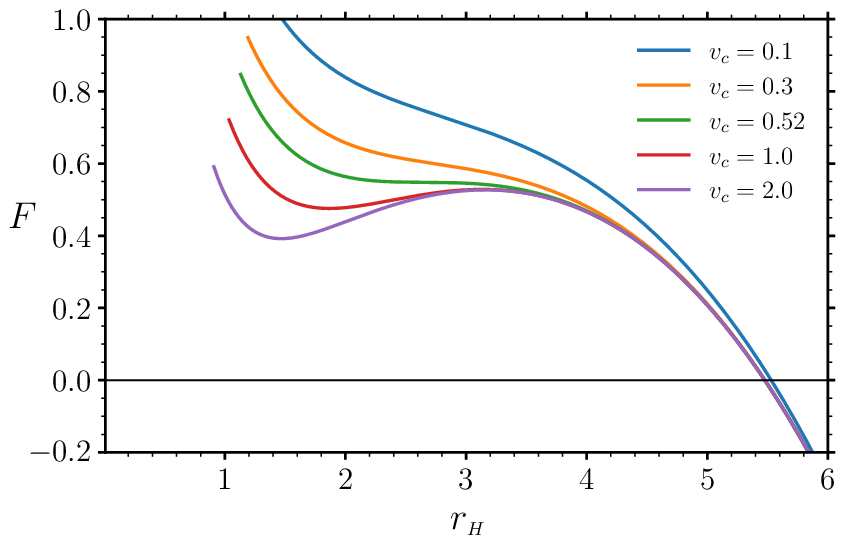}}
    \subfigure[$v_c=0.5$]{\includegraphics[width=0.49\textwidth]{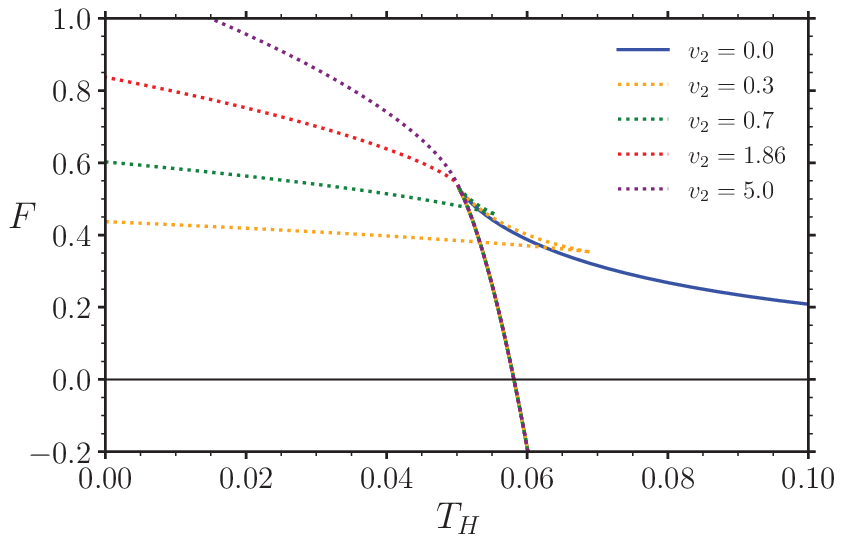}}
    \subfigure[$v_2=2.0$]{\includegraphics[width=0.49\textwidth]{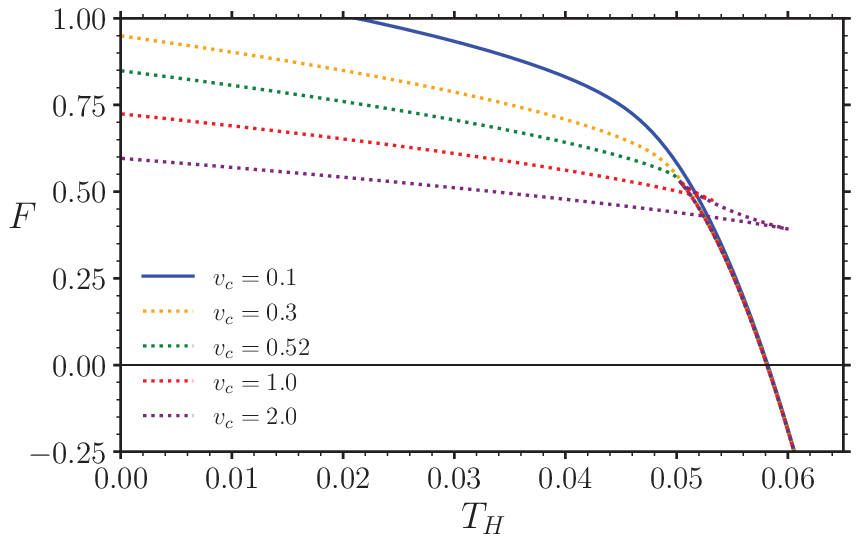}}
    \caption{Free energy $F$ versus horizon radius $r_H$ for $(a)$ and $(b)$, and $T_H$ for $(c)$ and $(d)$ with various parameter choices. }
    \label{fig:Fe01}
\end{figure}
Figure~\ref{fig:Fe01} shows the graph of the free energy $F$ as functions of $r_H$ and $T_H$ with $k=1$ and $\Lambda = -0.1$.
In Figs.~$(a)$ and $(b)$, the leftmost points on each curve indicate the positions of the event horizons of extremal black holes.
Figure $(a)$ shows the free energy as a function of $r_H$ for different values of $v_2$, where $v_c$ is $0.5$.
The blue curve where $v_2 =0$ represents the SAdS case.
The red curve where $v_2=1.86$ is the critical case, in which the curve includes the point $\frac{d^2 F}{dr^2_H}=0$.
This shows the case where no further phase transition occurs.
Figure $(b)$ shows the free energy as a function of $r_H$ for different values of $v_c$, where $v_2$ is $2.0$.
The set of critical parameters corresponds to $v_2=2.0\,, ~v_c=0.52$.
Figure $(c)$ shows the free energy as a function of $T_H$ for different values of $v_2$, where $v_c$ is $0.5$.
In this figure, we focus on points with identical free energy at the same temperature.
The blue curve represents the SAdS case, acting as our reference.
We consider the temperature at which this blue curve crosses the zero free energy point.
This point represents a phase transition between a system consisting of thermal particles in the AdS spacetime
and a black hole system in AdS spacetime at the same temperature, and vise versa.
A larger black hole in the AdS spacetime compared to this reference point has a lower free energy,
making it globally stable or more probable~\cite{Hawking:1982dh}.
Regarding the orange curve $(v_2 = 0.2)$, a triangular shape appears in the middle;
where the two lines intersect, the systems have the same temperature and free energy, allowing a phase transition.
In the critical case at $v_2 =1.86$, the triangular shape vanishes and no further first-order phase transition occurs;
this trend persists even at higher $v_2$ values.
Figure $(d)$ shows the free energy as a function of $T_H$ for different values of $v_c$, where $v_2$ is $2.0$.

Now, let us analyze Fig.~\ref{fig:Fe01} using Eq.~\eqref{freeeng}.
\begin{itemize}
\item Small $r_H$ (matter-dominated region)
\label{fre01}
\end{itemize}
In Eq.~\eqref{freeeng}, the last term, $-6v_2 \sqrt{2\pi v_c} \left( {\rm erf} \left(r_H \sqrt{\frac{v_c}{2}} \right) -1  \right)$,
is negligible across all regions. For small values of $r_H$, the second term, $\frac{15v_2 e^{-v_c r^2_H/2} }{r_H}$, dominates.
Since its value increases as $r_H$ decreases, its derivative is negative.

\begin{itemize}
\item Intermediate $(r_H)$ (mass-dominated region)
\label{fre02}
\end{itemize}
In the intermediate region of $r_H$, the first term, $3(2-k) r_H$, dominates.
As $r_H$ increases, this term also increases; therefore, its derivative is positive.

\begin{itemize}
\item Large $r_H$ (AdS-dominated region)
\label{fre03}
\end{itemize}
In the large $r_H$, the third term, $\Lambda r^3_H$, dominates.
Given that this term is negative, the free energy consequently becomes negative at large $r_H$.
However, in the region where the first and third terms compete,
the influence of the former can lead to positive free energy values.
Additionally, the derivative of the third term is negative.

\begin{figure}[H]
    \centering
    \subfigure[$v_2=0.3$]{\includegraphics[width=0.49\textwidth]{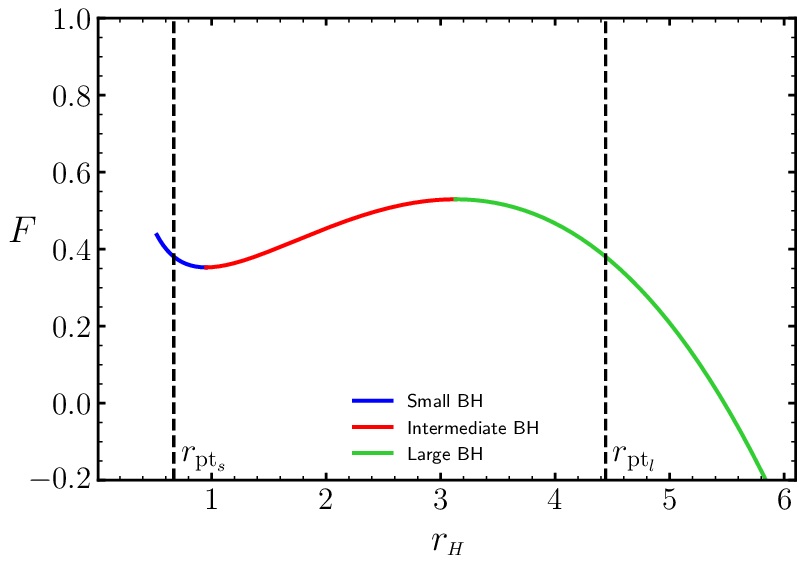}}
    \subfigure[$v_2=0.7$]{\includegraphics[width=0.49\textwidth]{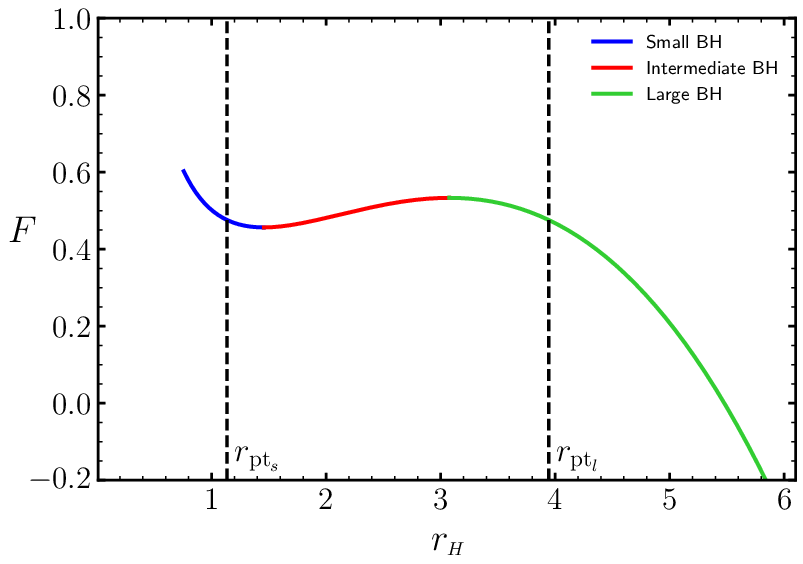}}\\
    \subfigure[$v_2=0.3$]{\includegraphics[width=0.49\textwidth]{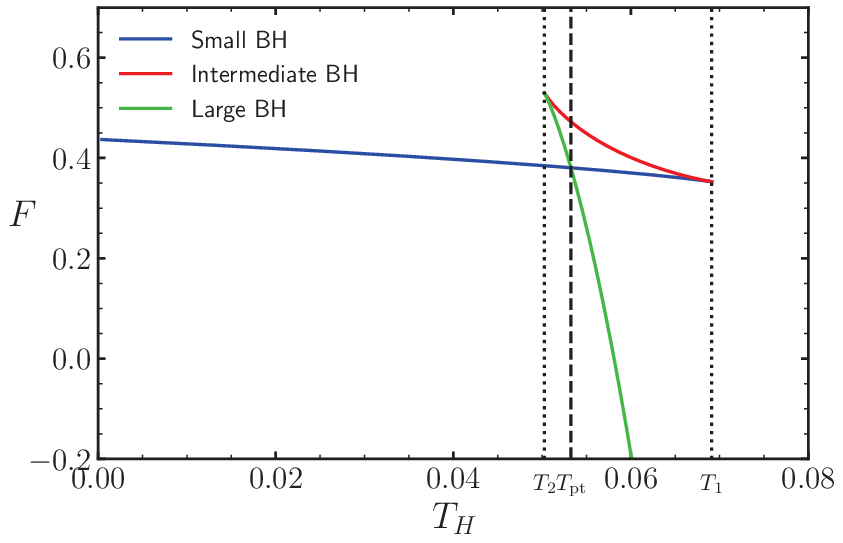}}
    \subfigure[$v_2=0.7$]{\includegraphics[width=0.49\textwidth]{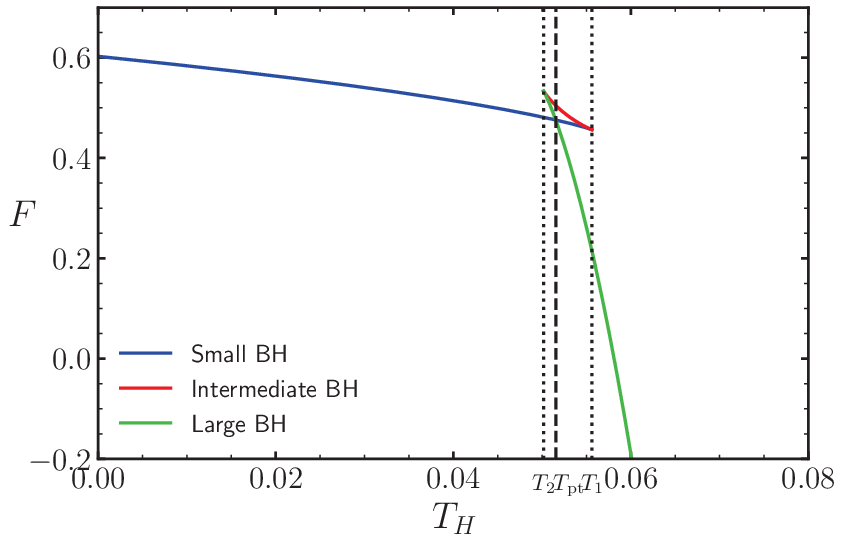}}\\
    \caption{A single typical free energy curve comprising three black hole phases as a function of $r_H$
              and temperature $T_H$.}
    \label{fig:FT2}
\end{figure}
Figure~\ref{fig:FT2} shows a single typical free energy curve, partitioned into three black hole phases, as a function of $r_H$ (top panels)
and temperature $T_H$ (bottom panels) for $v_2=0.3$ and $v_2=0.7$ with $v_c=0.5$ and $\Lambda=-0.1$.
In $(a)-(d)$, the blue, red, and green segments of the curves represent the small, intermediate, and large BH regions, respectively.
In $(c)$ and $(d)$, a phase transition between small and large BHs occurs at the intersection of the blue and green curves (temperature $T_{\rm pt}$).
The corresponding points at this temperature are labeled $r_{\rm pts}$ and $r_{\rm pts}$ in $(a)$ and $(b)$.
Furthermore, in $(c)$ and $(d)$, $T_1$, and $T_2$ denote the temperatures at which the blue/red and red/green curves intersect.
The points where the colors change are also identified in $(a)$ and $(b)$.

\section{Lyapunov exponent \label{sec4}}

\quad

We now analyze the Lyapunov exponent obtained from the particle's trajectory to determine its relationship with the black hole phase structure obtained in the previous Sec.~\ref{sec3}.

Thanks to the spherical symmetry of the spacetime under consideration,
the particle's orbit can be confined to the equatorial plane $(\theta = \pi/2)$ without loss of generality.
Furthermore, since this spacetime possesses a sufficient number of Killing vectors and Killing tensors,
the geodesic equations are both separable and integrable.
Focusing on null geodesics, the radial equation can be expressed as follows:
\begin{equation}
\label{effen}
\dot{r}^2 + V_{\mathrm{eff}}(r) = E^2 \,,
\qquad
V_{\mathrm{eff}}(r) = f(r)\,\frac{L^2}{r^2} \,,
\end{equation}
where the dot notation denotes the derivative with respect to an affine parameter $\sigma$,
$E$, and $L$ denote the conserved energy and angular momentum, respectively.
The location of the unstable local maximum of the potential is determined by
\begin{equation}
V'_{\mathrm{eff}}(r)|_{r=r_c} =0\,, \quad V''_{\mathrm{eff}}(r)|_{r=r_c} < 0 \,,
\end{equation}
where the double prime denotes the second derivative with respect to $r$.
This point coincides with the radius defining the photon sphere.
A positive Lyapunov exponent signifies divergence.
Following the definition used in Refs.~\cite{Cardoso:2008bp, Guo:2022kio}, it is expressed as follows:
\begin{equation}
\label{lyapunovgeneral}
\lambda = \sqrt{- \frac{r_c^2 f_cV_{\mathrm{eff}}''(r_c)}{2 L^2}}\,.
\end{equation}
Using the metric ansatz \eqref{stmetric}, this expression simplifies to
\begin{equation}
\label{lyapunovfinal}
\lambda = \sqrt{\frac{f(r_c)}{2 r_c^2} \left[ 2 f(r_c) - r_c^2 f''(r_c) \right]}.
\end{equation}
Equation~\eqref{lyapunovfinal} provides the Lyapunov exponent in terms of the metric function and its derivatives,
evaluated at the radius $r=r_c$ of the local maximum.

\begin{figure}
    \centering
    \subfigure[$v_2=0.3$]{\includegraphics[width=0.49\textwidth]{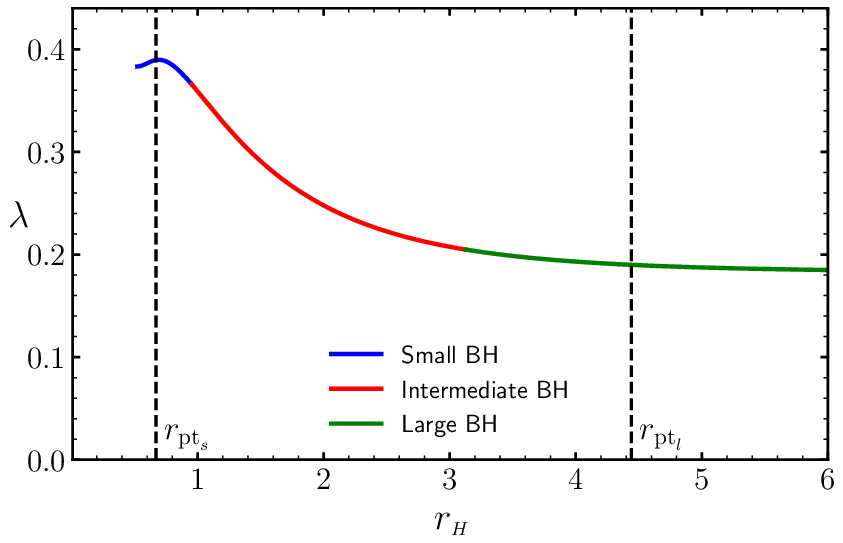}}
    \subfigure[$v_2=0.7$]{\includegraphics[width=0.49\textwidth]{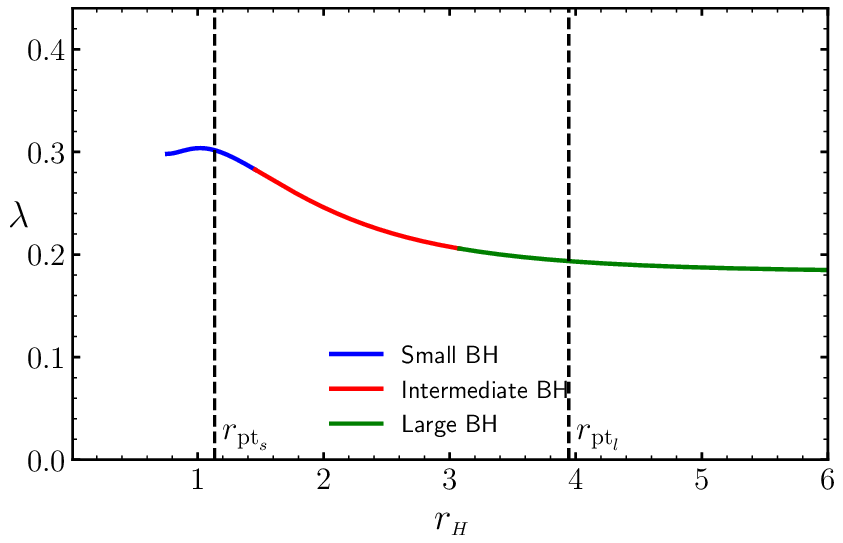}}
    \caption{Detailed branch structure of $\lambda(r_H)$ for two representative subcritical values.}
    \label{fig:FrH2}
\end{figure}
Figure~\ref{fig:FrH2} shows the plot of the Lyapunov exponent $\lambda$ as a function of $r_H$ with $v_c=0.5$ and $\Lambda=-0.1$.
Figures $(a)$ and $(b)$ present the Lyapunov exponent as a single typical curve, partitioned into three black hole phases.
Figure $(a)$ corresponds to $v_2=0.3$, and $(b)$ to $v_2=0.7$.
The endpoint of the blue curve on the left represents the case for the extremal black hole,
while the dashed black line indicates the phase transition point, $r_{\rm pt}$.
The Lyapunov exponent is large for the small black hole $(r_{\rm pt s})$ and small for the large black hole $(r_{\rm pt l})$.

\begin{figure}[H]
    \centering
    \subfigure[$v_2=0.3$]{\includegraphics[width=0.49\textwidth]{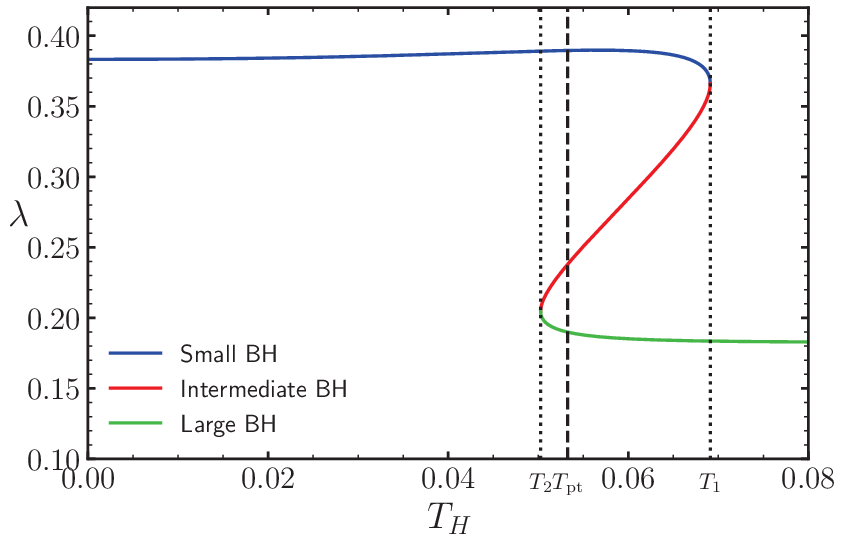}}
    \subfigure[$v_2=0.7$]{\includegraphics[width=0.49\textwidth]{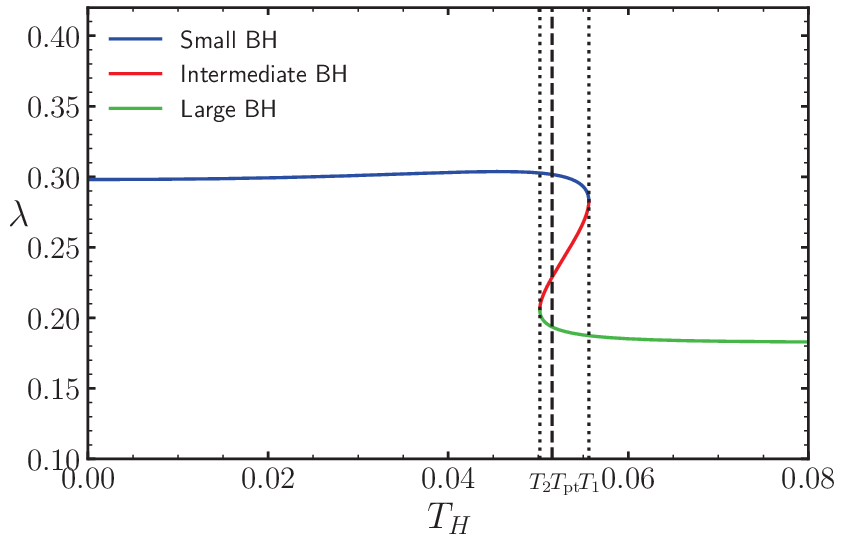}}\\
    \subfigure[$v_2=1.86$]{\includegraphics[width=0.49\textwidth]{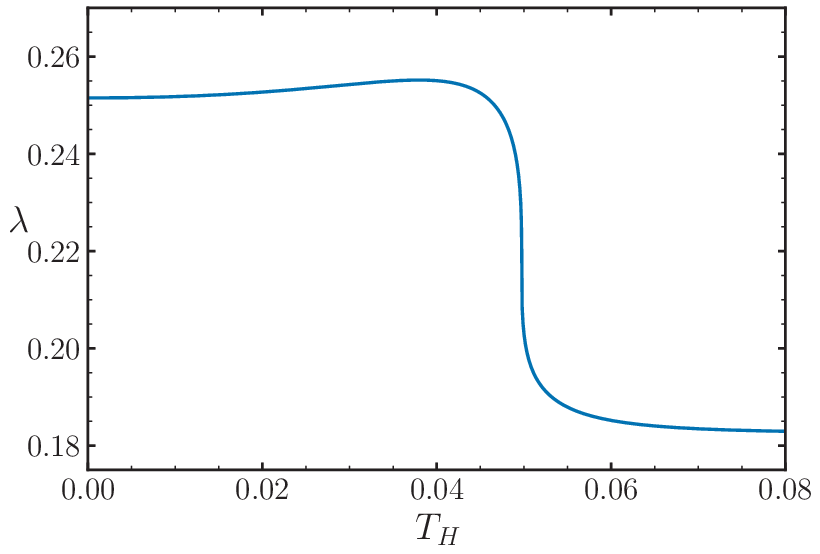}}
    \subfigure[$v_2=3.0$]{\includegraphics[width=0.49\textwidth]{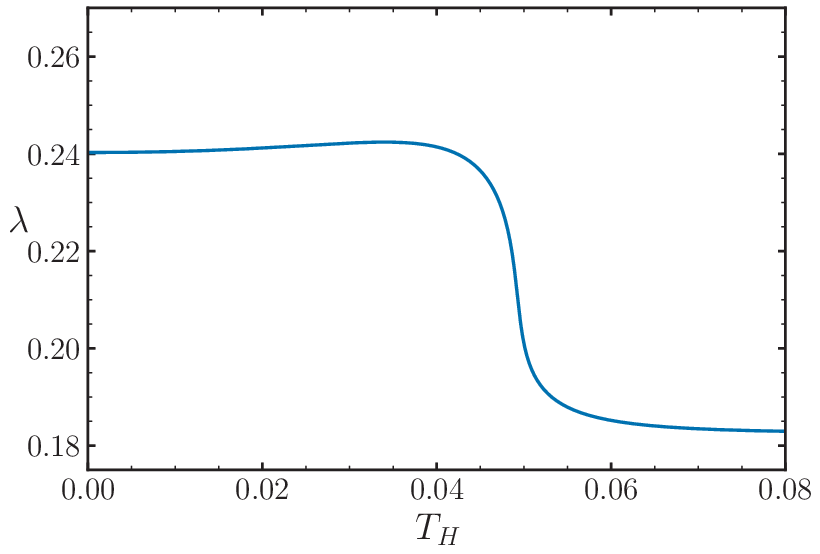}}
    \caption{Lyapunov exponent $\lambda$ versus the temperature $T_H$  for different values of $v_2$.}
    \label{fig:LT2}
\end{figure}
Figure~\ref{fig:LT2} shows the plot of the Lyapunov exponent $\lambda$ as a function of $T_H$ with $v_c=0.5$ and $\Lambda=-0.1$.
Figures $(a)$ and $(b)$ present the Lyapunov exponent versus temperature for the single typical free energy curve,
partitioned into three black hole phases (see Fig.~\ref{fig:FT2}).
We apply the methodology to black holes that coexist with a specific type of anisotropic matter.
This can be compared to Ref.~\cite{Guo:2022kio}, which was applied to RN-AdS black holes.
For small black holes at low temperatures and large black holes at high temperatures,
the Lyapunov exponent value exhibits a plateau, showing no significant variation.
Within the coexistence temperature range, the small, intermediate, and large BH regions possess different Lyapunov exponent values; notably, the exponent for the large BH phase is lower than that of the small BH phase.
The transition temperatures, $T_1$ (small to intermediate) and $T_2$ (intermediate to large), are also marked.
Figures $(c)$ and $(d)$ show the Lyapunov exponent as a function of temperature at critical values
$v_2$ and $v_c$, where first-order phase transitions vanish.

Critical exponents describe the behavior of physical quantities near a continuous phase transition.
They are considered universal; that is, they do not depend on the microscopic details of a physical system,
but rather on only a few general features of the system.
To quantitatively support the claim that $\lambda$ plays the role of an order parameter,
its critical exponent should be investigated.
For this purpose, we examine the difference of Lyapunov exponents at the phase transition point.
According to \cite{Lyu:2023sih, Gogoi:2024akv, Ali:2025ooh, Bezboruah:2025udi}, the critical exponent $\delta$ is related to the order parameter as
\begin{eqnarray}
\label{difflambd00}
\Delta \lambda = \lambda_s - \lambda_l \sim |T-T_c|^{\delta} \,.
\end{eqnarray}
We denote the Lyapunov exponent for small black holes as $\lambda_s$ and for large black holes as $\lambda_l$,
and $T_c$ represents the critical temperature. Near the critical point, the critical exponent that characterizes the change in the Lyapunov exponent was found to be $\frac{1}{2}$ and the ratio of the critical Lyapunov exponent $\lambda_c$ to $\Delta \lambda$ is given by
\begin{eqnarray}
\label{difflambd}
\frac{\Delta \lambda}{\lambda_c} &\simeq&\frac{\sqrt{T_{\rm pt}}}{\lambda_c} \left(\frac{\partial \Delta \lambda}{\partial r_H}\right)_c \left(\frac{1}{2} \frac{\partial^2 T}{\partial r^2_H} \right)^{-1/2}_c \sqrt{T_{\rm pt}/T_c -1} \nonumber \\
&=& 2.38276 \sqrt{T_{\rm pt}/T_c -1} \,,
\end{eqnarray}
where the number $2.38276$ is obtained numerically for our black hole system.

\begin{figure}[H]
    \centering
    \subfigure{\includegraphics[width=0.49\textwidth]{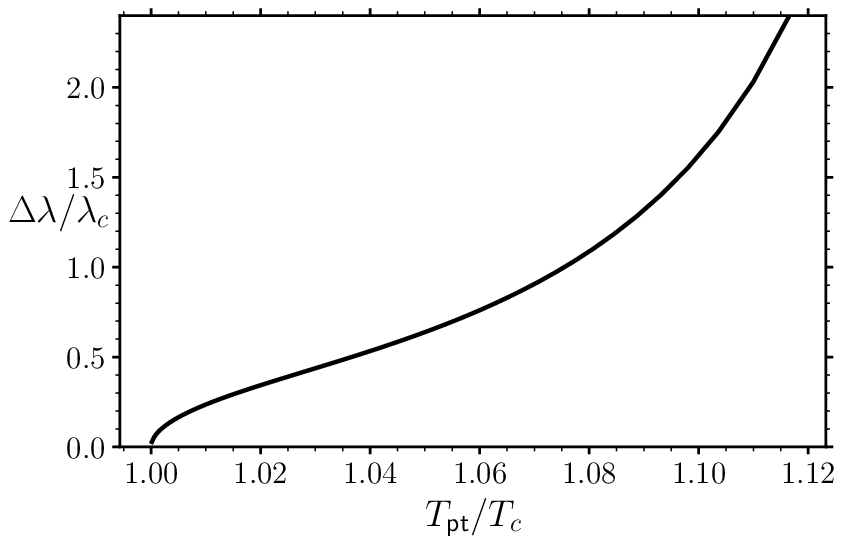}}
    \subfigure{\includegraphics[width=0.49\textwidth]{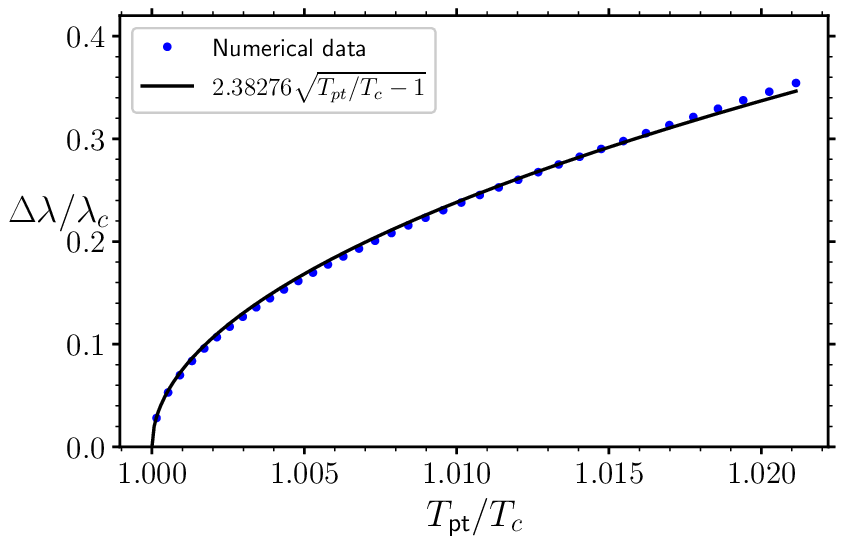}}
    \caption{$\Delta \lambda/\lambda_c$ vs $T_{\rm pt}/T_c$.}
    \label{critical}
\end{figure}
Figure~\ref{critical} shows the plot of $\Delta \lambda/\lambda_c$ vs $T_{\rm pt}/T_c$.
The left figure shows that the Lyapunov exponent remains non-zero at the phase transition.
As $T_{\rm pt}$ approaches $T_c$, it is demonstrated that $\Delta\lambda$ vanishes.
This behavior indicates that $\Delta\lambda$ could play the role of an order parameter.
The figure on the right displays the values near the critical temperature, where the blue dots represent the numerical data, and the black line corresponds to the theoretical values of Eq.~\eqref{difflambd}.

\section{ Summary and discussions \label{sec5}}
\quad

It is well established that dark energy and dark matter, the exact nature of which remains unknown,
comprise the vast majority of the energy budget of the Universe  in the current era.
In contrast, the ordinary matter we encounter makes up only a small fraction.
As a result, considerable research interest has transitioned from vacuum black hole solutions to the exploration of black hole solutions that coexist with matter fields.

As the connection between black hole thermodynamics and classical thermodynamics became established,
this field has become an active area of research for investigating various properties of black hole
thermodynamics, including its quantum nature.

In this situation, research investigating the physical properties of black holes coexisting with anisotropic matter is further motivated. We have newly constructed a black hole spacetime geometry coexisting with a specific form of anisotropic matter, assuming the presence of a cosmological constant $\Lambda$. Here, the cosmological constant includes both positive and negative values.
In this study, we are particularly interested in the case with a negative cosmological constant, where rich thermodynamic behavior emerges.

In Sec.~\ref{sec2}, we constructed a black hole that coexists with anisotropic matter in the presence of a cosmological constant. This one focuses on the case of a negative cosmological constant (AdS). In such a spacetime, the horizon structure is analyzed for all three possible horizon topologies: $k=1, 0, -1$. In Sec.~\ref{sec3}, we have focused on the $k=1$ case to analyze the black hole temperature and the heat capacity, which determines local stability. We also derived the Smarr relation and generalized the first law of black hole thermodynamics. Furthermore, we analyzed the free energy to determine the global stability.
The thermodynamic behavior is governed by the anisotropic matter parameters $v_2$ and $v_c$, which determine the existence of multiple thermodynamic branches. For suitable parameter ranges, the system exhibits a van der Waals-type phase transition between small and large black holes, whereas above the critical line only a single thermodynamic branch survives.
In Sec.~\ref{sec4}, the Lyapunov exponent is analyzed as an additional means to determine global stability.

We analyzed the thermodynamics and the phase transition of black holes coexisting with these matter fields, using the Schwarzschild-AdS (SAdS) case as a reference for comparison. The anisotropic matter exhibits thermodynamic properties similar to those of a Maxwell field. To analyze these properties, we adopted the canonical ensemble, assuming that the system maintains thermal equilibrium with its environment at temperature $T_H$, and the volume of the system is fixed. As a charged black hole system, we fixed the charge $v_2$ and $v_c$ rather than fixing the corresponding potentials. The corresponding free energy could be the Helmholtz free energy, which we analyzed as a function of temperature.

Let us summarize this from the perspective of the Helmholtz free energy.
We could consider the black hole as the ``system" and the AdS spacetime as the ``environment". Assuming that the system is in thermal equilibrium with the environment at temperature $T_H$, we would treat the mass $M$ as the energy required to create the black hole system. We did not consider the work ($pV$) required to create the space occupied by the system.
We would treat $T_H S$ as the heat that freely flows from the environment at temperature $T_H$ and could subtract that amount to obtain the free energy. A system naturally tends to minimize its Helmholtz free energy, which serves as a guiding principle in determining the global stability of the black hole geometries we are investigating.

Hawking and Page analyzed the phase transitions of black hole systems in AdS spacetime through the difference in free energy~\cite{Hawking:1982dh}, in which they studied the phase transition by treating thermal particles; which have the same temperature as the black hole system in AdS spacetime; as the system at the point where the free energy of the AdS black hole with the large-horizon  becomes zero. In the charged case, a transition from a small black hole to a large black hole becomes possible; in our case as well, a phase transition from a small black hole system to a large black hole system is allowed. Here, it is more appropriate to judge the phase transition in terms of global stability, i.e. which system is more probable. It would also be interesting to analyze the difference in Gibbs free energy in the grand canonical ensemble, where the work $pV$ required to create the space occupied by the system should be considered.

The Lyapunov exponent and the Poincar\'e section in phase space serve as important indicators for determining whether particle motion is chaotic within a curved spacetime. The Lyapunov exponent characterizes local chaotic behavior and measures sensitivity to initial conditions, while the Poincare section assesses global chaotic behavior. In particular, particle motion could display sensitivity to initial conditions even in integrable systems, highlighting the distinction between these two indicators within the Einstein theory of gravitation~\cite{Jeong:2023hom}. To evaluate the Lyapunov exponent, it is essential to first identify an unstable local maximum of the effective potential associated with a homoclinic orbit. Timelike geodesics associated with particles are expected to provide results that complement our null geodesic analysis. This investigation is deferred to future work.

Although the Lyapunov exponent does not directly determine the thermodynamic stability, both quantities originate from the same underlying spacetime geometry. The horizon radius determines the thermodynamic state through the Hawking temperature and free energy, while the same metric function simultaneously governs the effective potential of photon orbits and therefore their dynamical instability. Consequently, modifications of the spacetime geometry induced by anisotropic matter manifest themselves consistently in both thermodynamic and geodesic observable.

An order parameter could be defined as a physical quantity that indicates the degree of order in a thermodynamic system before and after a phase transition. Typically, this parameter would be zero in a disordered phase and non-zero in an ordered phase. It is important to note that the order parameter is not unique and there may be various choices.
In the Hawking-Page phase transition, the black hole radius acts as the order parameter.
References~\cite{Duan:1984ws, Wei:2021vdx, Jeon:2024yey, Zhu:2024zcl, Liu:2025iyl, Bao:2025wdp, Rani:2025mip, Yang:2025uul} employed Duan's topological current $\phi$-mapping theory
to define and utilize topological charges as order parameters.
Furthermore, analysis in Refs.~\cite{Guo:2022kio, Yang:2023hci, Shukla:2024tkw, Gogoi:2024akv, Gao:2025rep, Xie:2025auj} suggested that the Lyapunov exponent may also serve as an order parameter.
The identification and exploration of new order parameters in black hole phase transitions could prove to be quite intriguing~\cite{Yang:2025xck}.

\section*{Acknowledgments}
B.J. Ahmedov was supported by the National Natural Science Foundation of China (NSFC) under Grant No. U2541210. H.~Lee (NRF-2022R1I1A2063176), B.-H. Lee (RS-2026-25473640), W.~Lee (RS-2026-25484780), and CQUeST (RS-2020-NR049598) were supported by Basic Science Research Program through
the National Research Foundation of Korea funded by the Ministry of Education.
B.-H. Lee thanks  Asia Pacific Center for Theoretical Physics (APCTP) and
Korea Institute for Advanced Study (KIAS) for the hospitality during his visits, where a part of this project was done.
W.~Lee is grateful to Yun Soo Myung and Sang-Jin Sin for their helpful comments.

\end{document}